# Substitution of Lead with Tin Suppresses Ionic Transport in Halide Perovskite Optoelectronics

Krishanu Dey,[1] Dibyajyoti Ghosh,[2] Matthew Pilot,[3] Samuel R Pering,[4] Bart Roose,[5] Priyanka Deswal,[6] Satyaprasad P Senanayak,[7] Petra J Cameron,[3*] M Saiful Islam,[8*] Samuel D Stranks[1,5*]

[1]Cavendish Laboratory, University of Cambridge, Cambridge, UK

[2]Department of Materials Science and Engineering and Department of Chemistry, Indian Institute of Technology Delhi, Hauz Khas, India

[3]Department of Chemistry, University of Bath, Bath, UK

[4]Department of Materials, Loughborough University, Loughborough, UK

[5]Department of Chemical Engineering and Biotechnology, University of Cambridge, Cambridge, UK

[6]Department of Physics, Indian Institute of Technology Delhi, Hauz Khas, India

[7]Nanoelectronics and Device Physics Lab, National Institute of Science Education and Research, School of Physical Sciences, HBNI, Jatni, India

[8]Department of Materials, University of Oxford, Oxford, UK

Email for correspondence: sds65@cam.ac.uk, saiful.islam@materials.ox.ac.uk, chppjc@bath.ac.uk

## Abstract

Despite the rapid rise in the performance of a variety of perovskite optoelectronic devices with vertical charge transport, the effects of ion migration remain a common and longstanding Achilles' heel limiting the long-term operational stability of lead halide perovskite devices. However, there is still limited understanding of the impact of tin (Sn) substitution on the ion dynamics of lead (Pb) halide perovskites. Here, we employ scan-rate-dependent current-voltage measurements on Pb and mixed Pb-Sn perovskite solar cells to show that short circuit current losses at lower scan rates, which can be traced to the presence of mobile ions, are present in both kinds of perovskites. To understand the kinetics of ion migration, we carry out scan-rate-dependent hysteresis analyses and temperature-dependent impedance spectroscopy measurements, which demonstrate suppressed ion migration in Pb-Sn devices compared to their Pb-only analogues. By linking these experimental observations to first-principles calculations on mixed Pb-Sn perovskites, we reveal the key role played by Sn vacancies in increasing the iodide ion migration barrier due to local structural distortions. These results highlight the beneficial effect of Sn substitution in mitigating undesirable ion migration in halide perovskites, with potential implications for future device development.

## Introduction

Lead halide perovskites have shown a remarkable run in photovoltaic applications, with single junction solar cell efficiencies reaching close to 26% and tandem efficiencies (with Si) eclipsing 33%.[1] At the same time, external quantum efficiencies of perovskite LEDs have ascended to more than 20% for green, red and infrared emission.[2] In addition to their favourable optoelectronic properties, including high absorption coefficient,[3] large ambipolar carrier diffusion length[4] and facile bandgap tunability,[5] these developments in the device performance of halide perovskites have also been aided by the relative ease of processing of these materials using inexpensive solution-based methods that require low thermal budgets.[6] However, unlike more conventional semiconductors like silicon or III-V materials (e.g. GaAs), halide perovskites are 'soft' semiconductors where the constituent ions (i.e. $A^+$, $B^{2+}$ and $X^-$ ions in the standard $ABX_3$ stoichiometry) migrate in response to external stimuli, such as electrical voltage, temperature and light.[7-11] Such ionic transport is mediated by the presence of various point and extended defects that are inevitably formed during the low-temperature growth of polycrystalline films on non-epitaxial substrates,[12] and represents one of the biggest challenges that need to be tackled for demonstrating prolonged operational stability in solar cells and LEDs.

The manifestation of ion migration in lead (Pb) perovskite devices was first observed in the appearance of apparent large low-frequency dielectric constant and pronounced current-voltage hysteresis in perovskite solar cells.[13,14] Since then, a general consensus points towards the halide ions as the dominant mobile ionic species under standard operational conditions.[8, 15-17] These mobile ions have been shown to not only result in open circuit voltage gain and short circuit current loss under light soaking in perovskite solar cells,[18,19] but also affect the long-term performance of perovskite solar cells and LEDs.[20-23] In recent years, various strategies have been demonstrated to mitigate such effects of ion migration. For example, metal doping in Pb halide perovskites has been attempted as a mitigation measure for ion migration in operational devices.[24-28] Mixing differently sized organic cations (in the A-site of 3D perovskites and/or A' site of 2D perovskites) has also been another effective approach to hinder the ion transport and extend device operational stability.[29-32] In addition, modulating grain sizes as well as passivating defects along grain boundaries have all achieved promising results.[33-36] Similarly, interface engineering in Pb-based perovskite solar cells has also led to efficient suppression of ionic migration effects.[37-38]

Despite these efforts, there is still a limited understanding on the role of tin (Sn) substitution on the dynamics of ion migration in Pb halide perovskite optoelectronic devices. This is of importance given the recent fast pace developments in the fields of Sn and mixed Pb-Sn perovskite solar cells towards, among others, reduced lead and tandem solar cell applications.[39-41] Although the recent report from Ighodalo *et al.* seems to suggest negligible (or complete absence of) ion migration in pure-Sn perovskites,[42] their conclusions were based on lateral device structures and all-inorganic pure-bromide compositions, both of which are not directly relevant for hybrid perovskite-based optoelectronic applications with vertical charge transport (such as solar cells and LEDs). For example, such claims are found to be contradictory to the work from Thiesbrummel *et al.,*[19] where the presence of ionic migration effects was clearly evident in organic-inorganic mixed Pb-Sn iodide perovskite solar cells. Thus, it is vital to fully rationalize and understand the impact of Sn substitution on the dynamics of ion migration in Pb halide perovskite optoelectronic devices. In this work, using electrical measurements on solar cells, we show that Sn-containing Pb-based hybrid perovskites exhibit slower ionic diffusion when compared to their Pb-only analogues. With further insights obtained from first-principles calculations, we attribute these observations to the increased iodide migration barriers in Sn-containing perovskites due to the structural distortion associated with Sn vacancy defects. The key conclusions derived from this study are applicable to various Sn-containing perovskite optoelectronic devices and will help guide future materials and device development.

## Results & Discussion

### Material properties and device performance

The perovskite films and devices used in this study are based on solution-processed methylammonium (MA)-free compositions, viz. $FA_{0.15}Cs_{0.15}PbI_3$ (henceforth referred to as 'Pb perovskite') and $FA_{0.85}Cs_{0.15}Pb_{0.5}Sn_{0.5}I_3$ (henceforth referred to as 'Pb-Sn perovskite'), where FA refers to the formamidinium cation. X-ray diffraction (XRD) measurements performed on the films (**Figure 1**a) reflect the standard perovskite crystalline peaks observed for Pb and Pb-Sn perovskites, in addition to a small $PbI_2$ peak (at ~12.7°) appearing for the Pb perovskite. **Figure 1**b shows the absorption spectra of the corresponding films, demonstrating bandgaps ($E_g$) of 1.53 eV and 1.24 eV respectively for Pb and Pb-Sn perovskites. Photoluminescence (PL) spectra is also plotted alongside in **Figure 1**b, revealing peak emission at 827 nm (FWHM = 38 nm) and 1022 nm (FWHM = 80 nm) for the respective perovskites. Scanning electron microscopy (SEM) images of the films (**Figure S1**) yield grain sizes of 248 ± 102 nm and 308 ± 93 nm for Pb and Pb-Sn perovskites respectively.

We then fabricated Pb and Pb-Sn perovskite solar cells with p-i-n configuration with the architecture: ITO / [2-(9H-Carbazol-9-yl)ethyl] phosphonic Acid (2PACz) / perovskite / fullerene ($C_{60}$) / bathocuproine (BCP) / Cu. For Pb perovskites, JV scans (reverse and forward) of a characteristic device is displayed in **Figure 1**c, with open circuit voltage ($V_{oc}$) of 1.04 V, short circuit current density ($J_{sc}$) of 24.5 mA/cm², fill factor (FF) of 66.4% and power conversion efficiency (PCE) of 16.9%. The average photovoltaic parameters are summarized in **Table 1** (see **Figure S2** for statistical distributions). Moreover, the integrated $J_{sc}$ obtained from external quantum efficiency measurements (**Figure 1**d) is 23.4 mA/cm$^2$, which agrees well with the average $J_{sc}$ of 23.9 mA/cm$^2$ obtained from J-V measurements. We employed the more commonly used PEDOT:PSS as the hole transporting layer (HTL) for fabricating Pb-Sn perovskite solar cells (with an absorber thickness ~ 405 nm) as 2PACz is found to cause severe losses in FF of the corresponding devices (**Figure S3**). **Figure 1**e shows JV scans of a characteristic Pb-Sn perovskite solar cell, with the average of photovoltaic parameters given in **Table 1** (device statistics in **Figure S4**). The corresponding EQE spectrum is shown in **Figure 1**f, which gives an integrated $J_{sc}$ of 29.1 mA/cm$^2$ and agrees closely with that obtained from J-V measurements (28.8 mA/cm$^2$). While much thicker (> 800 nm) absorber layers are ideal for minimizing optical losses in Pb-Sn perovskite solar cells,[43] we did not obtain any appreciable improvement in the efficiency by increasing the concentration of perovskite solution from 1.35 M to 1.8 M (device statistics in **Figure S5**). We have intentionally not used any defect passivating additives in the perovskite solution or as post-deposition surface treatments in the fabricated device stacks because of their synergistic influence on ion migration. Therefore, conclusions derived from this study are applicable in general for various Sn-containing perovskite optoelectronic devices.

**Table 1. Average photovoltaic parameters of Pb and Pb-Sn perovskite solar cells.**

| Device | $V_{oc}$ (V) | $J_{sc}$ (mA/cm$^2$) | FF (%) | PCE (%) |
|---|---|---|---|---|
| Pb (2PACz) | 1.02 ± 0.04 | 23.9 ± 1.0 | 64.5 ± 2.2 | 15.6 ± 1.3 |
| Pb-Sn (PEDOT) | 0.77 ± 0.04 | 28.9 ± 0.9 | 67.8 ± 4.7 | 14.3 ± 1.1 |

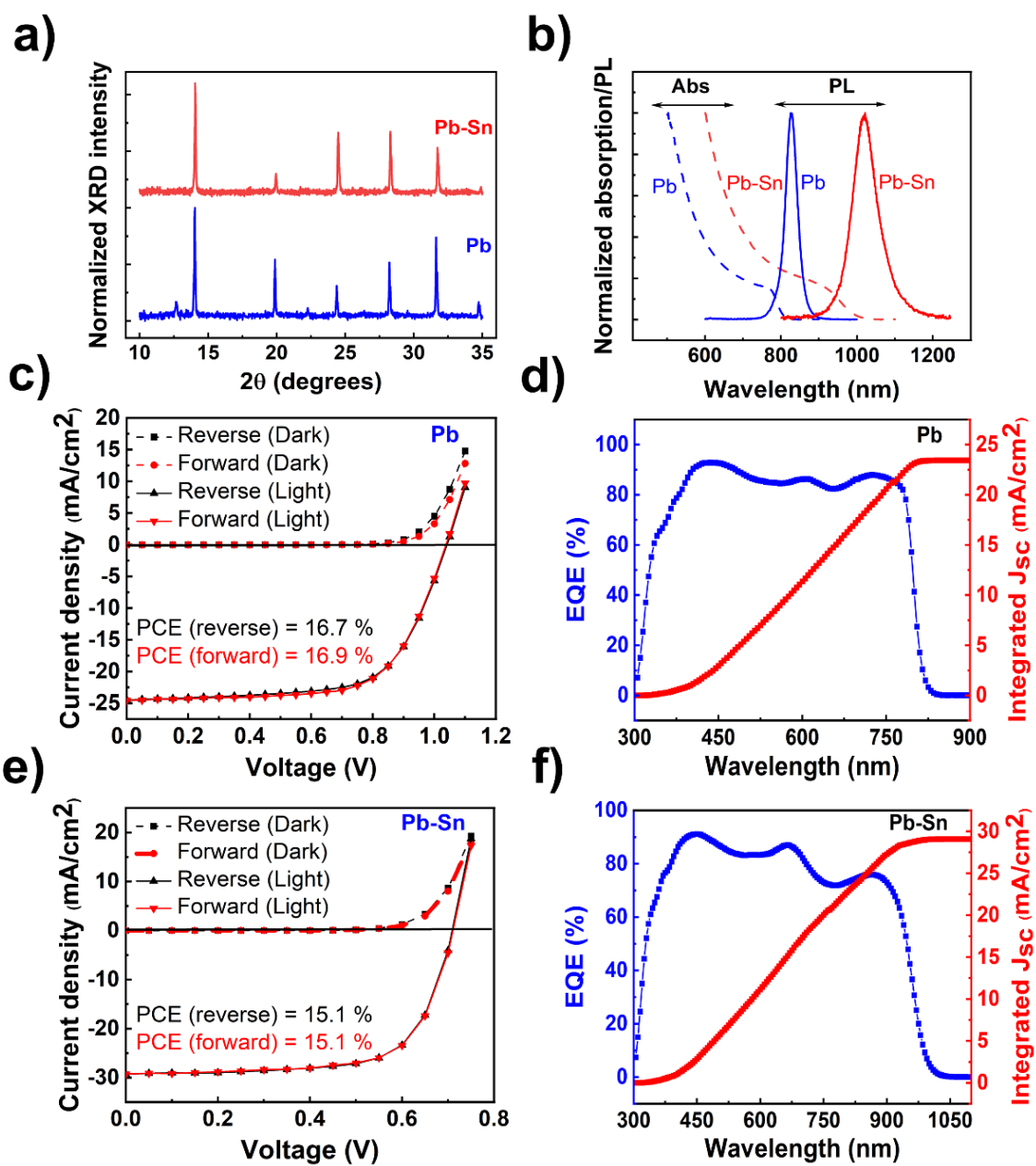

**Figure 1**: **Material and device characterization**. a) XRD patterns of Pb and Pb-Sn perovskite thin films. b) Absorption spectra (obtained from UV-visible-near-infrared spectroscopy) and photoluminescence spectra of Pb and Pb-Sn perovskite thin films. c) Light (under AM 1.5G illumination) and dark J-V scans of a characteristic Pb perovskite solar cell (with 2PACz HTL) with a scan rate of 100 mV/s. d) External quantum efficiency (EQE) spectra of the corresponding Pb perovskite solar cell. e) Light (under AM 1.5G illumination) and dark J-V scans of a characteristic Pb-Sn perovskite solar cell (with PEDOT:PSS HTL) with a scan rate of 100 mV/s. f) EQE spectra of the corresponding Pb-Sn perovskite solar cell. Here, 'Pb' refers to $FA_{0.85}Cs_{0.15}PbI_3$ and 'Pb-Sn' refers to $FA_{0.85}Cs_{0.15}Pb_{0.5}Sn_{0.5}I_3$. Optical bandgaps of the perovskite films were calculated from the absorption spectra using the Tauc-plot method. For PL measurements, a continuous-wave 405 nm laser was used as the excitation source. All the solar cell measurements were performed in air on encapsulated devices.

## Ionic transport properties from scan-rate dependent measurements

Having optimized material and device fabrication, we investigated the nature and extent of the role of ion transport in solar cells comprised of Pb and Pb-Sn perovskites. We performed J-V measurements over a range of scan rates from 5 mV/s to 1 V/s to capture any time-dependent performance changes in the devices. We did not lower the scan rate below 5 mV/s due to long measurement times (tens of minutes) and thus potential degradation of the devices due to bias stress, especially in dark conditions. Light J-V scans of Pb and Pb-Sn perovskite solar cells at three representative scan rates of 5 mV/s, 250 mV/s and 1 V/s are shown in **Figure 2**a-b (dark J-V scans in **Figure S6),** with the corresponding variation in the PCE of solar cells as a function of scan rates shown in **Figure S7**.

We display the variation of $J_{sc}$ as a function of scan rates in the Pb and Pb-Sn cells in **Figure 2**c-d, which shows an overall increase in $J_{sc}$ (during both reverse and forward scans) with increasing scan rates from 5 mV/s to 250 mV/s for both Pb and Pb-Sn perovskite solar cells. Such a trend can be rationalized by the fact that at slow scan rates, diffusing ions have enough time to react to the changes in voltage and hence these can move their equilibrium positions from the bulk (assuming roughly homogenous distribution at $V_{oc}$) towards the interfaces with the charge transport layers. Such movement of ions causes screening of the internal built-in field at short circuit, leading to a lowering of $J_{sc}$.[19] These observations point to the presence of mobile ions in both kinds of solar cells. Furthermore, a consistent increase in $V_{oc}$ is also observed by lowering the scan rates for the Pb device (**Figure 2**e), which may also originate from the prolonged light soaking effects on ion migration as observed by Herterich *et al.*[44] However, no such increase in $V_{oc}$ is observed for Pb-Sn devices.

Next, we calculated the variation of hysteresis index (HI) as a function of scan rates (**Figure 2**f) for both Pb and Pb-Sn perovskite solar cells, where HI in PCE is defined as HI (PCE) = $\frac{\text{PCE (reverse)} - \text{PCE (forward)}}{\text{PCE (reverse)}}$.[45] In the regime of scan rates < 50 mV/s, we observe a significant uptick in the HI (PCE) of Pb devices, while a relatively flat response is seen for their Pb-Sn analogues. Such a phenomenon of increasing hysteresis at lower scan rates for Pb perovskites is in agreement with reported drift-diffusion modelling on p-i-n perovskite solar cells involving ionic migration effects.[46-47]

Therefore, these observations indicate that the ionic migration in Pb-Sn devices is significantly slowed down as compared to that in Pb devices.

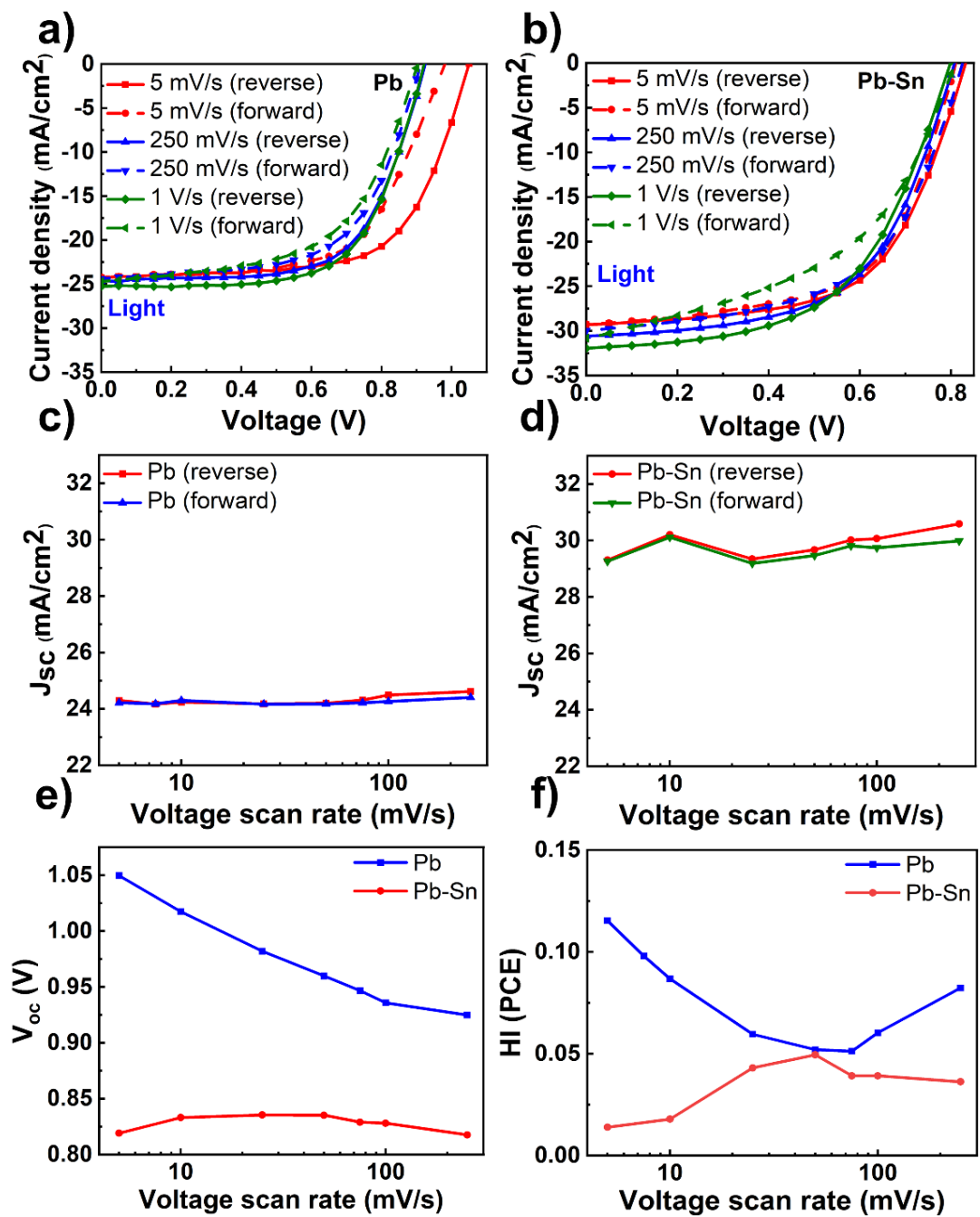

**Figure 2: Scan-rate dependent J-V measurements.** Light J-V scans of a) Pb and b) Pb-Sn perovskite solar cells at three scan rates: 5 mV/s, 250 mV/s and 1 V/s. c) $J_{sc}$ of Pb perovskite solar cells as a function of scan rates in the reverse and forward scan. d) $J_{sc}$ of Pb-Sn perovskite solar cells as a function of scan rates in the reverse and forward scan. e) Open circuit voltage of Pb and Pb-Sn perovskite solar cells as a function of scan rates. f) Hysteresis index (HI) in PCE of Pb and Pb-Sn perovskite solar cells as a function of scan rates.

## Ionic transport properties from impedance spectroscopy

To further understand the extent and impact of ionic diffusion, we employ impedance spectroscopy, which is a powerful technique to disentangle the ionic and electronic response in perovskite device stacks due to the characteristically different time scales of the two phenomena.[48–50] We carried out temperature-controlled impedance spectroscopy on Pb and Pb-Sn perovskite solar cells at open circuit under illumination with a blue 470-nm LED (**Figure 3**). To ensure the cells were stable during the measurements, they were loaded in a gas-tight device holder in a $N_2$ glovebox and also measured under nitrogen. Furthermore, to eliminate any complexity in the impedance response arising due to different HTLs, we used 2PACz as the common HTL for both Pb and Pb-Sn perovskite devices. Moreover, 2PACz-based devices were also found to be more stable under long-term measurements than the PEDOT-based devices.

**Figure 3**a shows the impedance spectra of a representative Pb perovskite solar cell measured at 25 °C over a frequency range of 100 mHz to 1 MHz. The Nyquist plot shows two-semi-circles, with a high frequency (hf) and lower frequency (lf) process clearly visible. The high frequency response gives a characteristic lifetime of 6.2 µs, which corresponds to a geometric capacitance ($C_{geo}$) of 26 nF and a recombination resistance ($R_{recomb}$) of 240 Ω. The time constant for the lower frequency process at 25 °C was 45 ms, which is consistent with the time constant we have previously measured for the 'lf' response attributed to ion diffusion inside perovskite crystallites.[30,51] Increasing the cell temperature from 21 °C to 40 °C results in an increase of characteristic frequencies for the 'lf' response (**Figure 3**b), indicating temperature-activated ionic transport in the devices. By fitting the low frequency time constants as a function of temperatures (**Figure S8**), an activation energy of 0.52 eV can be obtained which is in the range typically measured for iodide ion migration.[30]

In contrast, the impedance spectra of the Pb-Sn cell at 25 °C (**Figure 3**c) interestingly consists of only the 'hf' semicircle, with no evidence of the 'lf' ion mediated response. The 'hf' response has a characteristic lifetime of 5.8 µs, corresponding to a $R_{recomb}$ of 340 Ω and a $C_{geo}$ of 17 nF, which are similar to those obtained for the Pb cells. Moreover, even after increasing the cell temperature to 40 °C, it was still not possible to resolve any 'lf' response for the Pb-Sn device, even down to frequencies close to 1 mHz (**Figure 3**d). Therefore, these results indicate that ionic migration in these Pb-Sn cells is much slower than in the equivalent Pb perovskites. This suggests an activation energy for ion migration substantially higher than that obtained for our Pb perovskites (i.e. 0.52 eV) and we would likely have to go to higher temperatures or lower frequencies to see the response. This is not practical as typical perovskite cells are not stable for long periods at elevated temperatures or the very long measurement times needed to measure at frequencies below 1 mHz. In fact, extreme care needs to be taken when interpreting the low frequency response of all perovskite cells due to additional features that can be introduced by degradation.

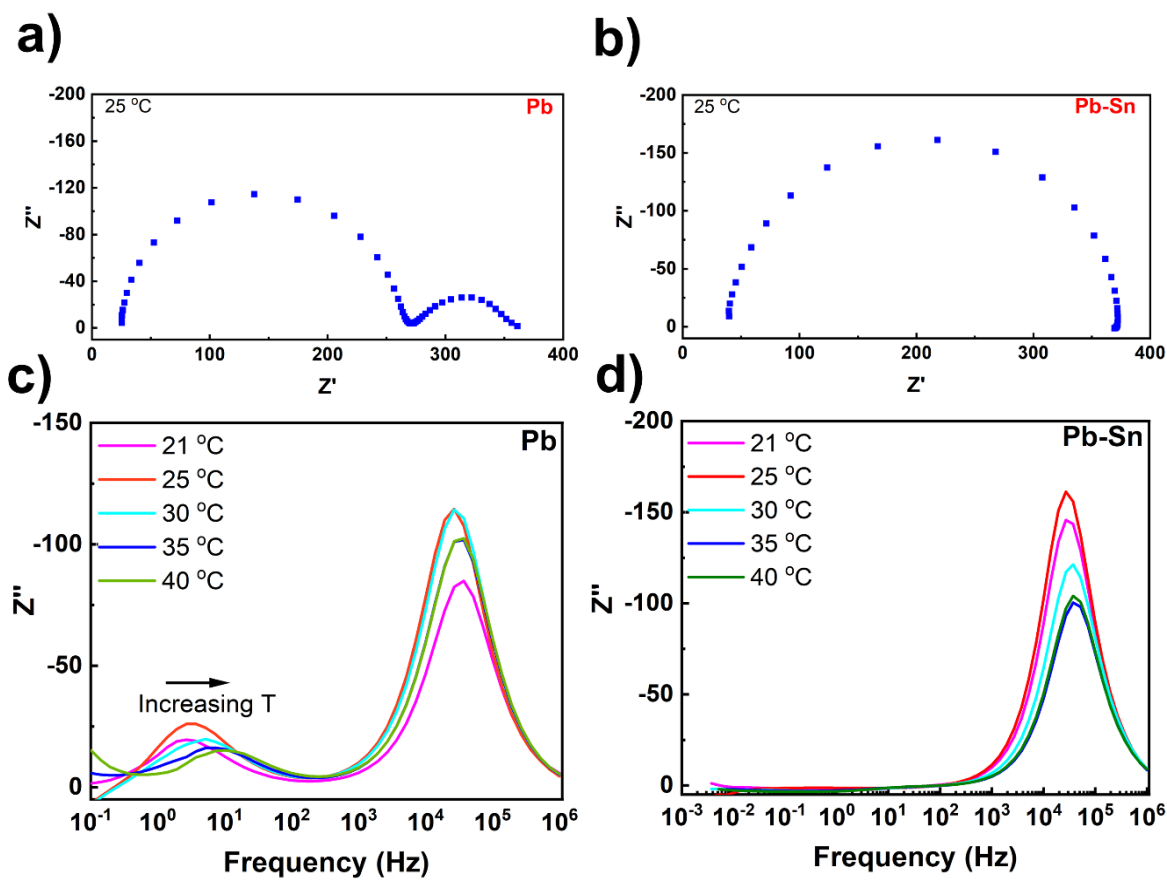

**Figure 3: Impedance spectroscopy measurements**. a) Nyquist plot of a Pb perovskite solar cell measured under illumination with a 470-nm LED at open circuit, while the cell was held at 25 °C throughout the measurement. b) Plots of the imaginary impedance against frequency for the corresponding Pb perovskite solar cell within a range of temperatures (21 – 40 °C). c) Nyquist plot of a Pb-Sn perovskite solar cell measured under illumination at open circuit, while the cell was held at 25 °C throughout the measurement. d) Plots of the imaginary impedance against frequency for the corresponding Pb-Sn perovskite solar cell within a range of temperatures (21 – 40 °C). All data were assessed through a Kramers-Kronig transformation which shows that the measured data are valid within these frequencies (**Figure S9**).

## Atomistic migration mechanisms through ab-initio simulations

To complement the experimental work with atomic-scale insights into ion migration mechanisms, we also performed systematic density functional theory (DFT) calculations (see details of the computational methodology in Methods). Previous studies by us and other groups have demonstrated that halide ions are the dominant migrating species via a vacancy mechanism in lead-halide perovskites.[8,17,30,52] Here, we investigate the migration pathways and energy barriers for iodide vacancy migration in Pb, mixed Pb-Sn and Sn perovskites.

To examine the impact of Sn substitution, we first focused on a $FAPbI_3$ perovskite structure, which allowed us to probe trends in ion migration energies in a systematic manner. Due to the tetragonal distortion in the lattice, there are two inequivalent iodide sites in $FAPbI_3$ and, consequently, we find that the two most probable pathways for iodide vacancy migration are equatorial-equatorial and axial-equatorial mechanisms, for which the calculated migration barriers are 0.34 eV and 0.45 eV, respectively (**Figure S10**). These results are in good agreement with previous experimental and computational work on Pb-based systems.[13,30] We anticipate the axial-equatorial iodide migration as the rate-determining step for long-range diffusion in the material, and focus on this pathway for the rest of the study. In addition, following similar methodology, our simulations on $FASnI_3$ find an activation energy of 0.36 eV for the axial-equatorial pathway of iodide migration.

We then explore iodide migration and associated energy barriers in the mixed-metal system $FAPb_{0.5}Sn_{0.5}I_3$. Due to such B-metal alloying, there are two inequivalent pathways for axial-equatorial migration involving Pb-centred and Sn-centred iodide diffusion (see **Figure S11**), for which we find energy barriers of 0.43 eV and 0.47 eV, respectively. Thus, our simulations suggest that B-metal alloying alone does not have any major impact on iodide migration barriers in these halide perovskites.

It is known that there is a significant population of Sn vacancy defects in Sn-containing perovskites due to their low formation energies.[53-55] Moreover, thermodynamic ionization levels of these defects lie close to the valence band maximum for mixed Pb-Sn perovskites, while they lie inside the valence band for pure-Sn perovskites. Therefore, these Sn vacancies can be easily ionized, thereby resulting often in the unintentional hole doping in Sn-containing perovskites.[53] If left uncontrolled, excessive hole doping can affect the short circuit current and fill factor of the fabricated solar cells and hence efforts are made to tune the growth conditions of the films by incorporating Sn-rich additives (such as $SnF_2$) in the precursor solutions to minimize the background doping levels in the perovskites.[56-57] Nevertheless, Sn vacancies still account for one of the primary sources for the inherent doping and $Sn^{2+}$ oxidation in Sn-containing perovskites and as such they affect the carrier recombination and transport properties of perovskites.[58] Therefore, we then investigated the impact of such Sn vacancies on the migration of iodide ions in $FAPb_{0.5}Sn_{0.5}I_3$ and $FASnI_3$. It is found that the most stable $SnI_2$ Schottky-type defect comprises of the Sn vacancy and two I vacancies ($V_{Sn}$, $2V_I$) at adjacent sites rather than at well separated positions (**Figure S12**). With this Schottky defect cluster ($V_{Sn}$, $2V_I$), we examined three distinct iodide ion migration pathways labelled A, B, and C (more details in Supplementary Note 3 in Supporting Information) in **Figure 4**a, and their energy profiles are shown in **Figure 4**b. For $FAPb_{0.5}Sn_{0.5}I_3$, these A, B and C pathways involve different sequences of equatorial-axial type hops leading to energy barriers greater than 0.9 eV and to a rate-limiting ion migration energy for long-range diffusion of 1.45 eV. Following similar migration paths in $FASnI_3$, we again find energy barriers for iodide ion migration greater than 0.9 eV and to a rate-limiting migration energy for long-range diffusion of 1.12 eV. (**Figure S13**). **Table 2** summarises the rate-limiting migration barriers for iodide ions in the different model perovskite systems with and without the presence of Sn vacancies. Thus, our simulations clearly suggest that iodide migration near Sn vacancy defects in $FAPb_{0.5}Sn_{0.5}I_3$ and $FASnI_3$ face high energy barriers (> 1.1 eV) compared to <0.5 eV for Pb perovskites due to severe local structural distortion. Overall, we highlight the important role of Sn vacancies in arresting the migration of iodide ions in Sn-containing perovskites, in good accordance with our experimental observations of much slower ion diffusion in mixed Pb-Sn devices.

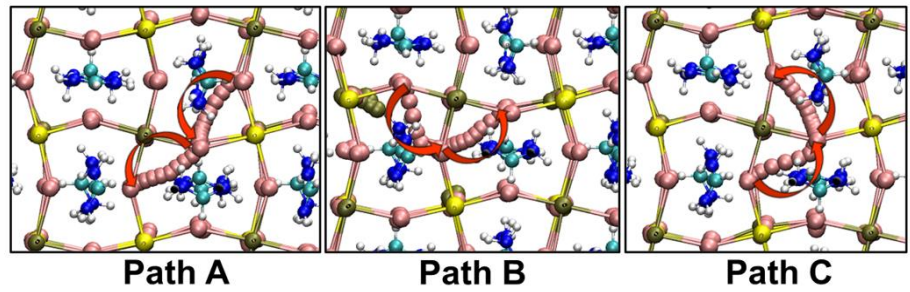

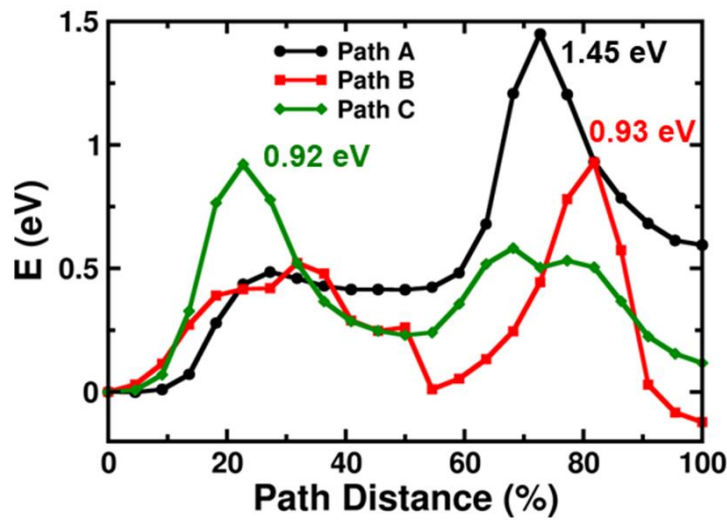

**Figure 4**: **Ion migration pathways through atomistic ab initio simulations**. a) Inequivalent iodide migration paths (A, B and C) in FAPb$_{0.5}$Sn$_{0.5}$I$_3$ involving the SnI$_2$ Schottky-type defect comprised of a Sn vacancy and two I vacancies. Here, brown, olive, pink, turquoise, blue and white spheres represent lead, tin, iodine, carbon, nitrogen and hydrogen atoms respectively. The red arrows point to the direction of iodide migration in the lattice. b) Energy profiles for iodide ion migration via paths A, B and C in FAPb$_{0.5}$Sn$_{0.5}$I$_3$.

**Table 2. Rate-limiting ion migration energies for long-range iodide ion transport in different model perovskite systems.**

| Perovskite system | Ion migration energy (eV) |
|---|---|
| FAPbI$_3$ | 0.45 |
| FASnI$_3$ (without Sn vacancies) | 0.36 |
| FASnI$_3$ (in presence of Sn vacancies) | 1.12 |
| FAPb$_{0.5}$Sn$_{0.5}$I$_3$ (without Sn vacancies) | 0.47 |
| FAPb$_{0.5}$Sn$_{0.5}$I$_3$ (in presence of Sn vacancies) | 1.45 |

# Conclusion

The impact of substituting Pb with Sn on the ion migration properties of halide perovskite optoelectronic devices has been investigated using a combination of experimental and computational techniques. Short circuit current loss obtained at lower scan rates (< 50 mV/s) indicates the prevalence of ion migration in both Pb and Pb-Sn perovskite solar cells, but the kinetics of ion transport are suppressed in mixed Pb-Sn systems as inferred from scan-rate dependent hysteresis measurements. These results are further corroborated by temperature-dependent impedance spectroscopy measurements performed on the fabricated devices at open circuit and under light illumination, which suggest a substantial lowering of ionic diffusion with the partial substitution of Pb with Sn. In addition, atomistic ab initio simulations highlight the role of Sn vacancies in increasing the iodide ion migration barriers (>1.1 eV) in Sn-containing perovskites due to severe local structural distortion, which corroborates and rationalises our experimental observations of much slower ion diffusion in mixed Pb-Sn perovskite solar cells. Overall, our findings can be generalized for a variety of Pb halide perovskite optoelectronic devices, where the benefit of Sn substitution in suppressing ionic migration effects may lead to enhanced operational stability and improved device architectures.

# Acknowledgements

K.D. acknowledges the support of the Cambridge Trust for the Cambridge India Ramanujan Scholarship and Cambridge Philosophical Society for the research studentship. This work received funding from the European Research Council under the European Union's Horizon 2020 research and innovation programme (HYPERION, grant agreement no. 756962). M.S.I. and D.G acknowledge ARCHER2 supercomputer resources via membership of the HEC Materials Chemistry Consortium funded by the EPSRC (EP/R029431). D.G. acknowledges the IIT Delhi SEED Grant (PLN12/04MS), the Science and Engineering Research Board (SERB), Department of Science and Technology (DST), India for Start-up Research Grant SRG/2022/00I234 and the IIT Delhi HPC facility for computational resources. P.J.C. and M.P. thank the EPSRC Centre for Doctoral Training in New and Sustainable Photovoltaics (EP/L01551X/2) for funding. S.P.S acknowledges funding support from

SERB (SRG/2020/001641 and IPA/2021/000096). S.D.S. acknowledges support from the Royal Society and Tata Group (UF150033). The authors acknowledge the Engineering and Physical Sciences Research Council (EPSRC) for funding (EP/R023980/1, EP/T02030X/1, EP/V012932/1). For the purpose of open access, the authors have applied a Creative Commons Attribution (CC BY) licence to any Author Accepted Manuscript version arising from this submission.

## Contributions

K.D. conceived the idea and designed the experimental plan with supervision from S.D.S, M.S.I. and P.J.C.; K.D. optimized the processing of perovskite films and transport layers and accordingly fabricated all the devices for electrical characterization. K.D. performed and analyzed UV-Vis, XRD and PL measurements on the perovskite films. K.D. also carried out and analyzed standard and scan-rate dependent J-V measurements on perovskite solar cells. D.G. conducted DFT calculations on the migration barrier of iodides in different perovskite systems with assistance from P.D. and interpreted the results with M.S.I; M.P. and S.R.P measured impedance spectroscopy on perovskite solar cells and analyzed the data with P.J.C.; B.R. measured the top-view SEM of the perovskite films. S.P.S. provided useful inputs on the understanding of ionic migration in perovskite devices. K.D. interpreted all the data and wrote the first draft of the manuscript with detailed contributions from all the authors.

## Competing Interests

S.D.S. is a cofounder of Swift Solar. The remaining authors declare no competing interests.